# Production of ultra-low radioactivity NaI(Tl) crystals for Dark Matter detectors

Authors:   Y. Zhu (1), S.H. Yue (1), Z.W. Ge* (1), Y.W. Zhu (1), X.J. Yin (1), I. Dafinei (2), G. DImperio (2), M. Diemoz (2), V. Pettinacci (2), S. Nisi (3), C. Tomei (2), H.B. Zhao (1), B. Xu (1), J. Fang (1), Q.W. Tu (1)

[1]Shanghai Institute of Ceramics, Chinese Academy of Sciences, 200050, Shanghai, P.R.China
[2]INFN Sezione di Roma, Piazzale A. Moro 2, 00185 Roma, Italy
[3]INFN Gran Sasso National Laboratory, Via G. Acitelli, 22, 60100 L'Aquila, Italy

*Corresponding author: Z.W.Ge; Shanghai Institute of Ceramics, Chinese Academy of Sciences, 200050, Shanghai, P.R.China; Tel.: +862169987886; e-mail: gezengwei@mail.sic.ac.cn

*Abstract*
*Scintillating NaI(Tl) crystals are widely used in a large variety of experimental applications. However, for the use as Dark Matter (DM) detectors, such crystals demand a high level of radio-purity, not achievable by means of standard industrial techniques. One of the main difficulties comes from the presence of potassium that always accompanies sodium in alkali halides. On the other hand, the arguable DM detection by DAMA experiment using NaI(Tl) scintillating crystals requires a reliable verification able to either confirm the existence of DM or rule out the DAMA claim. Ultra-low radioactivity NaI(Tl) crystals, particularly with very low potassium content, are therefore indispensable to overcome the current stalemate in Dark Matter searches. Nonetheless, apart from DAMA-LIBRA experiments, to date, no other experiment has succeeded in building a detector from NaI(Tl) crystals with potassium content of ppb level. This work describes recent results in the preparation of ultra-radio-pure NaI(Tl) crystals using a modified Bridgman method. A double-walled platinum crucible technique has been designed and reliability tests show that 5 ppb of potassium in the NaI(Tl) crystals of 2 and 3 inches in diameter can be achieved starting from NaI powder with potassium content of the order of 10 ppb. The potassium excess is segregated in the tail-side of the as grown ingot where measured potassium concentration is above 20 ppb. The purifying effect of Bridgman growth for larger NaI(Tl) crystals is currently being tested. The work also reports on scintillation parameters of our NaI(Tl) crystals measured in a dedicated setup conceived for naked, hygroscopic crystals. The reproducible and reliable production of ultra-low radioactivity NaI(Tl) crystals reported in this work will hopefully spur the construction of new DM search experiments and, anyway, clarify the controversial DAMA-LIBRA results.*

## 1. Introduction

Due to their remarkable scintillating properties, NaI(Tl) crystals are the most widely used inorganic scintillators. Nevertheless, the use of these crystals is not obvious when particular characteristics are requested for particle physics and astrophysics experiments. In the case of Dark Matter (DM) detectors, an extremely high level of radio-purity is compulsory, which is not achievable by means of standard industrial techniques. The DAMA/LIBRA [1] experiment based on 250 kg of radio-pure NaI(Tl) crystal scintillators is the only one which claims to detect DM using the direct detection approach. The experiment reports an annual modulation of the number of detection events, caused by the variation of the velocity of the detector relative to the DM halo in our galaxy.

Several experiments based on NaI(Tl) detectors were dedicated to clarifying the DAMA/LIBRA claims: KIMS [2], SABRE [3], ANAIS [4], DM-Ice [5], PICO-LON [6], COSINE [7]. The main challenge for all these experiments consists in building their detector out of ultra-radio-pure NaI(Tl) crystals. Given their chemical affinity, NaI crystals typically are contaminated significantly with 40K which may undergo an electron capture that is accompanied by the emission of Auger electrons of 3.2 keV and 3.0 keV and constitutes the main background source in the region of interest (ROI) for DM search. The success of DM experiments based on NaI(Tl) crystals relies therefore on reducing the concentration of radioactive contaminants at least down to the level of DAMA/LIBRA crystals (K < 20 ppb and U, Th of the order of µBq/kg). This proved to be a very challenging technological task, which remained unmatched so far. The current work describes the Bridgman growth of such NaI(Tl) crystals with a double-walled platinum crucible technique. The motivation behind it, is the desire to make available reproducible and reliable ultra-low radioactivity NaI(Tl) crystals thus offering the possibility to build competitive DM experiments or improve the performance of current ones and, finally, clarify the controversial DAMA/LIBRA results.

## 2. Material requirements and crystal growth

### 2.1. Overview

The obtainment of a radio-pure crystal is a process in which all steps from the selection and procurement of the raw material, the treatment to prepare the powder for crystal growth and the crystal growth itself have to be carried out in a coherent way respecting strict measures to avoid the introduction of any kind of contamination. The production process of NaI(Tl) crystals for DM use is divided into three major phases: NaI powder synthesis, crystal growth and cutting/polishing of the scintillator element.

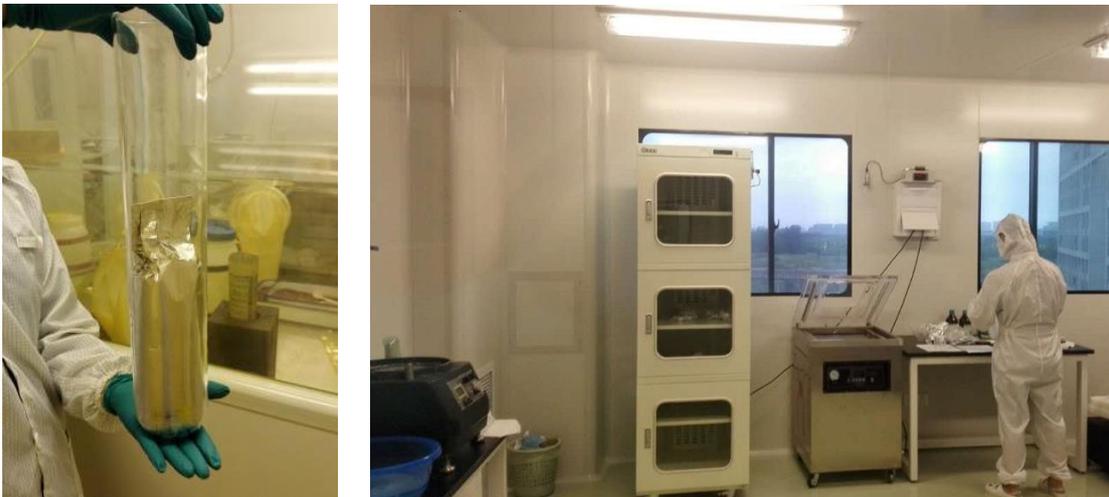

Fig. 1 Crystal production at SICCAS: Double walled Platinum crucibles for crystal growth (left) and a dry-clean room with facilities for NaI(Tl) crystals' mechanical processing.

### 2.2. NaI(Tl) crystal growth

For the crystal growth performed at SICCAS (Fig. 1), a double-walled platinum crucible technique was designed, taking into account two aspects:

    1) platinum held at SICCAS is of a high purity and was successfully used for the production of high-purity $TeO_2$ crystals for CUORE experiment [8];

    2) the use of thin platinum foils is hazardous because of the risk of melt leakage during crystal growth. In addition, careful preparation of crucible charge was ensured in order to avoid the chemical reaction

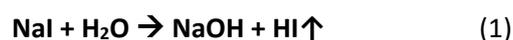

$$NaI + H_2O \rightarrow NaOH + HI\uparrow \qquad (1)$$

and consequent crucible damaging by the gaseous HI. The double-walled platinum crucibles are filled with ultrapure NaI(Tl) powder, sealed and placed into alumina refractory tubes which at their turn are placed into modified Bridgman furnaces [9]. Crucibles are heated to about 680°C and kept at this temperature for 4 hours after which the growth process is driven by lowering the crucible at a rate of 0.6 mm/h and raising the furnace temperature by about 0.5 °C/h. At the end of the growth process the furnaces are cooled down to room temperature slowly (typically 48 h, depending on the crystal dimensions) in order to avoid crystal cracks caused by thermal stress. Scintillating elements are further made by cutting and shaping in standard conditions followed by a final polishing made in a ultra-dry clean room using consumables certified for their radio-purity[8]

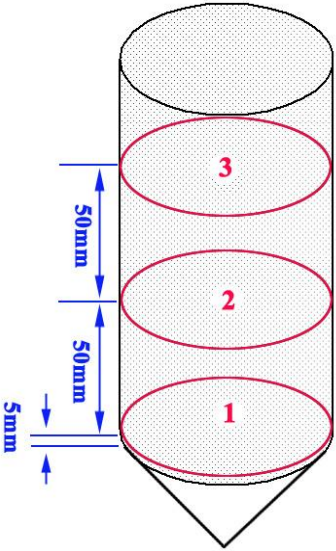

Fig. 2 Sampling positions for NaI and NaI(Tl) crystals

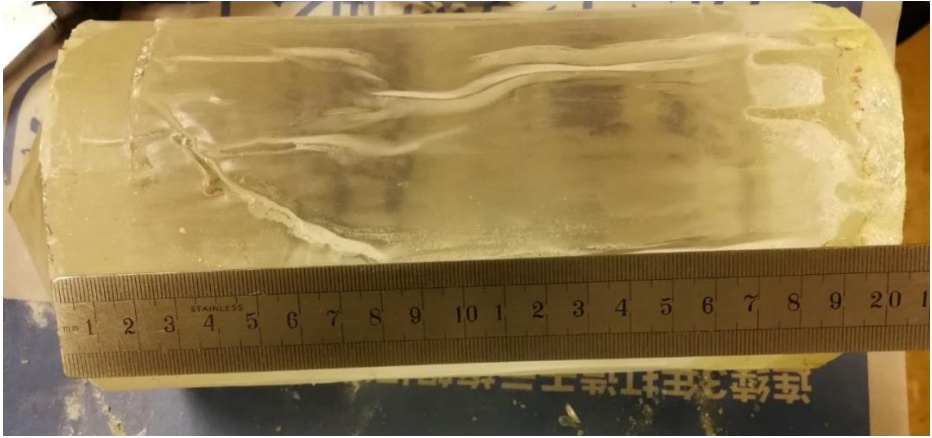

Fig. 3 NaI(Tl) ingot Φ4", L8" grown in this work

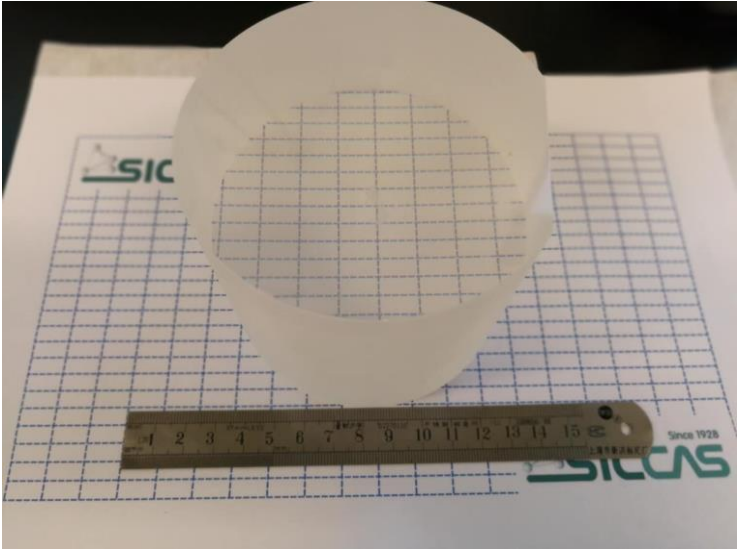

Fig. 4 NaI(Tl) crystal Φ4", L4" prepared in this work

## 3. Experimental techniques and results

### 3.1. Radio-purity measurements

The key issue of any experiment aimed at validating the DAMA claim is the radio-purity of NaI(Tl) scintillating crystal. The detector radio-purity requirements are very demanding and, due to the chemical affinity, the potassium is the most dangerous contaminant in the NaI crystal. In this work, the monitoring of the radio-purity (particularly the potassium concentration) for materials and consumables was made in two laboratories (at INFN and SICCAS) by mass spectrometry (ICP-MS). Following improvements made to the instruments and measurement protocols, detection limits of the order of few $10^{-9}$ g/g have been reached for the potassium detection in NaI matrix.

Detailed data related to radio-purity of raw materials and grown crystals are listed in Tab. 1.

Table 1. ICP-MS results of potassium

| Item | Size | *Position | Potassium (ppb) |
|---|---|---|---|
| NaI powder | - | - | 11 |
| TlI powder | - | - | 220 |
| NaI crystal | Φ2"x4" | 1 | 3 |
| NaI crystal | Φ2"x4" | 2 | 5 |
| NaI crystal | Φ2"x4" | 3 | 12 |
| NaI crystal | Φ3"x4" | 1 | 5 |
| NaI crystal | Φ3"x4" | 3 | 21 |
| NaI(Tl) crystal | Φ3"x4" | 1 | 6** |
| NaI(Tl) crystal | Φ3"x4" | 3 | 22** |

*The sampling positions are shown in Fig.2
**The doping level of Tl was 0.2wt% in the powder

As shown in Table 1, reliable tests given the fact that ~5 ppb of potassium in the NaI and NaI(Tl) crystals of 2" and 3" in diameter can be achieved starting from NaI powder with potassium content of 11 ppb. The potassium excess is segregated in the tail-side of the as grown ingot where measured potassium concentration is above 20 ppb for 3" crystals.

The growth of a large NaI(Tl) ingot (Φ4", L8" as shown in Fig. 3) and the mechanical processing of a large NaI(Tl) crystal (Φ4", L4" as shown in Fig. 4) were succeeded and the radio-purity measurements will be reported in a future work.

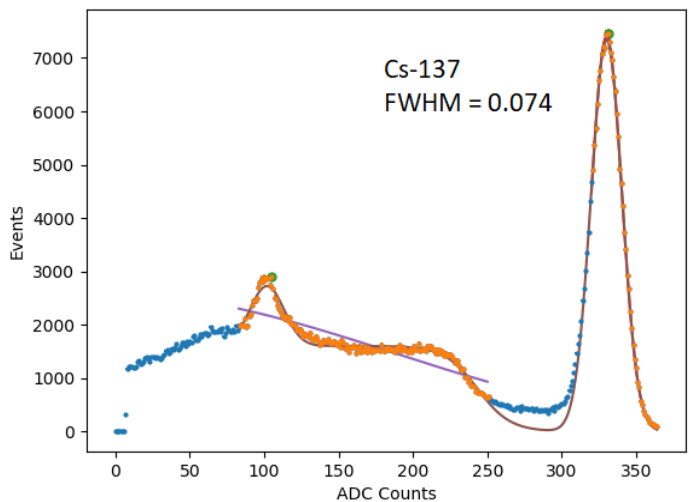

Fig. 5 Typical γ–spectrum ($^{137}$Cs source) for a NaI(Tl) crystal produced at SICCAS (Φ3", naked, no optical contact with PM)

### 3.2. Light Yield (LY) and Energy Resolution (ER) measurement

Light yield (LY) of a scintillating crystal is closely related to its energy resolution (ER). High light yield of the scintillator and high efficiency of the light detector is thus fundamental to have good energy resolution at low energy, where dark matter (DM) signal is expected.

LY and ER measurement is made in a dedicated setup designed to measure hygroscopic, naked crystals of up to Φ4" and 8" length. Fig.5 gives a typical gamma spectrum analyzed with a dedicated *python*™ script.

### 4. Concluding remarks and future work

Background is the key issue for DM experiments. Radio-purity, specially the reduced potassium content of NaI(Tl) crystals was a challenging task successfully solved at SICCAS. The reliable production of ultra-low radioactivity NaI(Tl) at SICCAS will hopefully spur the development of new experiments to confirm or discard DAMA/LIBRA claim. We will conduct more crystal growth tests, including the growth of larger mass crystals (7kg), for which dedicated furnaces are already commissioned. We will also improve the LY setup and reconstruction algorithm to achieve better resolution.

**References**
**[1]** R. Bernabei et al., **doi:10.1140/epjc/s10052-013-2648-7**
**[2]** P. Adhikari et al., **Eur. Phys. J. C 76 (2016) 185**
**[3]** E. Shields et al., **Phys. Proc. 61(2015) 169-178**
**[4]** J. Amaré et al., **Eur. Phys. J. C 76 (2016) 429**
**[5]** *DM-Ice collaboration*, E. Barbosa de Souza et al., **Phys. Rev. D 95 (2017) 032006**
**[6]** K. Fushimi et al., **J. of Phys.: Conf. Ser. 718 (2016) 042022**
**[7]** R. Maruyama, **COSINE-100, talk at the TAUP 17**
**[8]** C. Arnaboldi et al, **J. Cryst. Growth 312 (2010) 2999-3008**
**[9]** Jiayue Xu, Shiji Fan, Baoliang Lu, **J. Cryst. Growth 264 (2004) 260-265**